\documentclass[preprint,referee]{aastex}
\usepackage{epsfig,amsmath,amsfonts,amssymb,graphicx,subfigure,enumitem,color}
\usepackage{amssymb}
\usepackage{amsmath}
\usepackage[varg]{txfonts}
\usepackage{natbib}
\usepackage{url}
\usepackage{graphicx} 
\usepackage{epstopdf}
\graphicspath{{./media/abhi/SolankiR/Blowout_Jet/Revision/Apj_Arxiv_Format/aastexv6.2/V62/APj/}}
\usepackage{anyfontsize}
\usepackage{float}
\usepackage{textcomp}
\usepackage{color}
\usepackage{hyperref}

\shorttitle{R. Solanki \textit{et al.}}
\shortauthors{Twin CME Launched by a Quiet-Sun Blowout Jet}
\begin{document}
\title{Twin CME Launched by a Blowout Jet Originated from the Eruption of a Quiet-Sun Mini-filament }
\email{ritikas.rs.phy15@itbhu.ac.in}
\author{R. Solanki}
\affil{Department of Physics, Indian Institute of Technology (BHU), Varanasi - 221005, India}
\author{A. K. Srivastava}
\affil{Department of Physics, Indian Institute of Technology (BHU), Varanasi - 221005, India}
\author{Y. K. Rao}
\affil{Department of Physics, Indian Institute of Technology (BHU), Varanasi - 221005, India}
\author{B. N. Dwivedi}
\affil{Department of Physics, Indian Institute of Technology (BHU), Varanasi - 221005, India}
\begin{abstract}
We study a quiet-Sun blowout jet which is observed on 2014 May 16 by the instruments on board \textit{Solar Dynamics Observatory} (SDO). We find the twin CME as jet-like and bubble-like CME simultaneously as observed by LASCO C2 on board \textit{Solar and Heliospheric Observatory} (SoHO), \textit{Solar Terrestrial Relation Observatory} (STEREO\_A and STEREO\_B/COR2). They are respectively associated with the eruption of the northern and southern sections of the filament. A circular filament is rooted at the internetwork region at the base of the blowout jet. The collective magnetic cancellation is observed by SDO/HMI line of sight (LOS) magnetograms at the northern end of the filament, which makes this filament unstable and further makes it to erupt in two different stages. In the first stage, northern section of the filament is ejected, and causes an evolution of the northern part of the blowout jet. This part of the blowout jet is further extended as a collimated plasma beam to form a jet-like CME. We also observe the plasma blobs at the northern edge of the blowout jet resulting from Kelvin-Helmholtz (K-H) instability in its twisted magneto-plasma spire. In the second stage, southern section of the filament erupts in form of deformed/twisted magnetic flux rope which forms the southern part of the blowout jet. This eruption is most likely caused by the eruption of the northern section of filament, which removes the confined magnetic field of the southern section of the filament. Alternative scenarios may be a magnetic implosion between these magnetic structures confined in a much larger magnetic domain. This eruption of southern section of the filament further results in the bubble-like CME in the outer corona.
\end{abstract}
\section{Introduction} \label{sec:intro}
Solar coronal jets are the magnetically driven confined plasma eruptions which may be rooted in the lower solar atmosphere and further evolved in the solar corona. These dynamics occur in all solar environments such as quiet Sun, active region and coronal holes (e.g., Shimojo et al.1996, Nistico et al. 2009, Panesar et al. 2016, Raouafi et al. 2016). Solar coronal jets can be divided in two classes based on the temperature range of the ejected plasma as i) hot jets and ii) cool jets. In the hot jets, plasma is ejected into the solar corona along the open magnetic field lines as seen in UV, EUV, and X-ray wavelengths (e.g., Shibata et al. 1992, Shibata et al. 2007, Shen et al. 2012, Sterling et al. 2015, Liu et al. 2015, Shen et al. 2017). The temperature range of the ejected plasma is $\approx 4\times10^{6}$ K for solar X-ray jets and $\approx 10^{5}$ K for EUV jets (e.g., Shimojo et al. 1996, Shimojo et al. 1998). Jets observed in H$\alpha$ wavelength are termed as cool jets which are known as solar surges having the typical temperature $\approx 10^{4}$ K (e.g., Shibata et al. 1992, Yokoyama et al. 1995, Jiang et al. 2007, Uddin et al. 2012, Kayshap et al. 2013). 
Howerver cool coronal jets are also observed whose plasma is maintained at chromosphere, TR temperatues (e.g., Srivastava et al. 2011, Kayshap et al. 2013).
Solar coronal jets have their width, which ranges from $5 \times 10^3$ $km$ to $1 \times 10^5$ $km$, the height ranges from $\approx 10^{4}$ km to $4\times10^{5}$ km, and the velocity ranges from $10$ $km sec^{-1}$ to $1000$ $km sec^{-1}$ (e.g., Shimojo et al. 1996, Nistico et al. 2009). The width of the jets can be determined by the length of the emerging bipoles as illustrated by Shen et al. 2011.\par
According to the magnetic topology of the coronal jets, they can be categorized in these following types- i) Eiffel Tower (ET) jets  ii) $\lambda$ jets (e.g., Nistico et al. 2009) etc. In ET jets, photospheric bipolar magnetic field reconnects with the ambient open unipolar magnetic field of opposite polarity at the top of its loop. In the $\lambda$-shaped jets, the bipolar magnetic field reconnects with opposite polarity open magnetic field at their footpoints. Solar jets are classified as the anemone jets and two-sided loop jets based on their morphological structures and triggering mechanisms (e.g., Shibata et al. 1994, Yokoyama et al. 1995, Tian et al. 2017).
There is another classification of solar coronal jets according to the standard model of the jet (Shibata et al. 1992). The jets which follow the standard model of the jets are called the standard jets and the others which do not follow such characteristics are termed as blowout jets. The concept of the blowout jets is firstly introduced by Moore et al. 2010 using Hinode/XRT observation of an X-ray jet. Blowout jets possess broad spire and bright magnetic base arch, while standard jets have narrow spire and dim magnetic base arch. In the blowout jet, the core field of the magnetic base arch carries the cool plasma ($\approx 10^{4}$-$10^{5}$ K) filaments. In the blowout jet, there is enough twist and shear in the base arch field, so when base arch field and ambient field reconnect at the current sheet, this sheared and twisted arch field is also erupted. On the contrary, in the standard jet, the base arch is inactive and does not participate in the eruption (e.g., Moore et al. 2010, Moore et al. 2013, Sterling et al. 2015). It should be noted that the blowout jet involves two times reconnection, while the standard jet have only one time reconnection. Sometimes, the eruption of base-arch field takes the large form of the eruption and drags CMEs (chen et al. 2011). \par
Hong et al. 2011 presented the observation of blowout jet in quiet Sun region where mini-filament is ejected during eruption and associated with a mini coronal mass ejections (CMEs). Shen et al. 2012 observed an active region blowout jet which is associated with two CMEs as one is bubble- like CME and another is jet-like CME, where bubble like CME relates cool component of the jet and jet-like CME relates hot componet of the jet. Pucci et al. 2013 have done the comparative study of the blowout jet and standard jet and found that blowout jets have ten times higher magnetic energy as compared to the standard jets. Adams et al. 2014 have analyzed the blow-out jet event which originates from the on-disk coronal hole and shows the different characteristics. \par
Miao et al. 2018 have analyzed a coronal blowout jet eruption which is associated with a EUV wave at its leading top and complex CME structures (jet-like and bubble-like CME) where filament eruption is observed during the blowout jet. Shen et al. 2018 have found the close relationship between the coronal jets and the EUV waves at different scales (spatial and temporal) in their observational event, where these EUV waves were propelled by the coronal jets. The observational results of Shen et al. 2018 have shown the presence of the EUV waves along with the coronal jets, where these EUV waves are generated by the lateral expansion of loop system due to the coronal jet eruption. Shen et al. 2018 have discussed the observations of arc-shaped EUV waves, a quasi-periodic fast propagating (QFP) wave, and a kink wave simultaneously with an active region coronal jet eruption.    
Zhu et al. 2017 have observed an active region blowout jet and investigated the 3D magnetic structure of the blowout jet and found that the kink instability is a possible triggering mechanism for this blowout jet. There are many research articles which deal with these interesting eruptive events and describe the different possible mechanisms of these eruptions (e.g., Liu et al. 2008, Murawski et al. 2011, Chen et al. 2015, Liu et al. 2015, Alzate et al. 2016). Wang et al. 1998 showed that EUV jets may be directly extended into the form of white-light jet-like CMEs. Coronal jets can cause multiple CMEs through interaction with remote structures (Jiang et al. 2008) or through the self-evolution of coronal blowout jets (Shen et al. 2012, Miao et al. 2018). So far, studies on this issue are still very scarce.\par
In this paper, we have studied a blowout jet eruption observed on 2014 May, 16 which is evolved due to the eruptions of the various segments of a quiescent filament. Here we observed the twin CME generation with this blowout jet. The jet-like CME is associated with the northern part of the blowout jet eruption and the bubble-like CME is driven from the eruption of the southern part of the blowout jet. Both the parts of the blowout jet are associated with successive eruptions of the various segments of a filament. The K-H unstable blobs are also observed in the northern part of the blowout jet on its spire. Observational data and its analyses are described in Section 2. In Section 3, we illustrate the observational results and driving mechanisms of the observed blowout jet, and the kinematics of the twin CME. In the last Section, discussion and conclusions are presented.
\par
\section{Analysis of Observational Data}
\subsection{Observations from \textit{Solar Dynamics Observatory} (SDO)/\textit{Atmospheric Imaging Assembly} (AIA)}\par
We use high temporal and spatial \textit{Solar Dynamics Observatory} (SDO; Pesnell et al. 2012) data for the multi-wavelength study of the blowout jet. \textit{Atmospheric Imaging Assembly} (AIA; Lemen et al. 2012) observes full disk Sun in transition region and coronal emissions upto 0.5 R$_\odot$  above the solar limb. SDO/AIA provides the full disk images of the Sun in three UV wavelength bands 1600 \AA~, 1700 \AA~, 4500 \AA~ and in seven EUV wavelength bands 304 \AA~, 171 \AA~, 193 \AA~, 211 \AA~, 335 \AA~, 131 \AA~, 94 \AA~ covering the temperature range from 0.6 MK to 16 MK with $1.5^{\prime\prime}$ spatial resolution, and $0.6^{\prime\prime}$ pixel width. SDO/AIA captures the full-disk images of the Sun with 12 s. cadence in EUV filters and 24 s. in UV filters. We have taken the SDO/AIA data on 2014 May, 16 during the time period of 03:30:00 UT - 05:10:00 UT for the selected region of  $500^{\prime\prime}$ to $900^{\prime\prime}$ in the X-direction and $-300^{\prime\prime}$ to $0^{\prime\prime}$ in the Y-direction. We have downloaded SDO/AIA data from Joint Science Operation Center (JSOC) \footnote{\url{http://jsoc.stanford.edu.}}. Standard subroutines of SSWIDL (Freeland et al. 1998) are used for aligning and scaling AIA images as observed in different filters.\par 
\subsection{Observations from \textit{Solar Dynamics Observatory} (SDO)/\textit{Helioseismic Magnetic Imager} (HMI)}\par
We use \textit{Helioseismic Magnetic Imager} (HMI; Scherrer et al. 2012) data to examine the morphology and topology of magnetic field at the footpoints of the observed blowout jet. SDO/HMI provides the full-disk line-of-sight (LOS) magnetic flux in the Fe\textsc{i} 6173 \AA~ spectral line. It has 45-second temporal resolution, $0.5^{\prime\prime}$ pixel width and $1^{\prime\prime}$ spatial resolution. We have analyzed HMI magnetograms for the time period of 02:59:24 UT - 05:30:54 UT. SDO/HMI data has been rotated and aligned with the SDO/AIA data by using the standard subroutines of SSWIDL.\par
\subsection{Observations from \textit{Global Oscillation Network Group} (GONG)}\par
We use GONG H$\alpha$ (Harvey et al. 2011) data for study the dynamics of the filament eruption. We have downloaded H$\alpha$  data from GONG data archive \footnote{\url{https://gong.nso.edu/.}}. This gives the full disk H$\alpha$ data with 1 minute cadence and $1^{\prime\prime}$ spatial resolution in 6563 \AA~ wavelength.\par
\subsection{Observations from \textit{Solar and Heliospheric Observatory} (SoHO)/\textit{Large Angle and Spectrometric Coronagraph} (LASCO)}\par
We use LASCO CMEs data obtained from CME catalogue, which is available in CDA website \footnote{\url{http://cdaw.gsfc.nasa.gov/CME_list.}}. \textit{Large Angle and Spectrometric Coronagraph} (LASCO; Brueckner et al. 1995) on board the \textit{Solar and Heliospheric Observatory} (SoHO; Domingo et al. 1995) daily identify the coronal mass ejections (CMEs) in the images of the solar corona since 1996. LASCO has three telescopes named C1, C2, C3. LASCO observes the white light images of the solar corona from 1.1 R$_\odot$ to 30 R$_\odot$. LASCO C2 coronagraph images the solar corona from 1.5 R$_\odot$ to 6 R$_\odot$, while C3 coronagraph images the solar corona from 3.5 R$_\odot$ to 30 $R_\odot$. We use LASCO C2 and C3 data for detailed scientific investigation of narrow CME which is associated with some part of the blowout jet that further erupted in the outer corona after its origin into the quiet-Sun.\par
\subsection{Observations from \textit{Solar Terrestrial Relation Observatory} (STEREO)/\textit{Sun Earth Connection Coronal and Heliospheric Investigation} (SECCHI)}\par
We use \textit{Sun Earth Connection Coronal and Heliospheric Investigation} (SECCHI; Howard et al. 2008) on board STEREO-A and STEREO-B spacecraft data to analyse the kinematics of the blowout jet, jet-like and bubble-like CMEs. We use the images of the \textit{Extreme Ultraviolet Imager} (EUVI; Wuelser et al. 2004) of SECCHI for determining the kinematics of the blowout jet, and the COR2 data of SECCHI for kinematics of jet-like and bubble-like CMEs. The field of view (FOV) of EUVI and COR2 is 1-1.7 $R_\odot$ and 2.5-15 $R_\odot$ respectively. We have downloaded the SECCHI data from the UKSSDC-STEREO archive \footnote{\url{https://www.ukssdc.ac.uk/solar/stereo/data.html.}}.\par
\section{Observational Results}
\subsection{Source Location of the Blowout Jet Evolved due to Quiet-Sun Filament Eruption}\par  
The blowout jet, which we have studied in the present paper was observed in the quiet Sun region on 2014, May 16. This quiet Sun region is near the western side of the NOAA AR12058 (S11W40), and lies in the fourth quadrant of the solar disk co-ordinates. The location of the blowout jet is Xcen = $660^{\prime\prime}$ and Ycen = $-150^{\prime\prime}$, and the initiation time is about 04:08:43 UT. This blowout jet is evolved due to the multiple stages of filament eruption which is rooted in the internetwork region of the quiet-Sun. We will describe physical picture of such unique plasma dynamics in the coming sub-sections.  \par 
The multi-wavelength behaviour of the blowout jet is seen in the composite image of different SDO/AIA filters (\textit{cf.} upper panel of Figure 1). We have plotted the composite image of blowout jet and surrounding regions in AIA 1600 \AA~, AIA 304 \AA~ and in HMI line of sight (LOS) magnetogram at 04:11 UT. This image collectively shows the behaviour of the blowout jet simultaneously in the emissions from UV continnumm to the transition region, and also the magnetic field polarites around its footpoint. The chosen field of view (FOV) for this composite image is $400^{\prime\prime} \times 300^{\prime\prime}$ as $500^{\prime\prime}$ to $900^{\prime\prime}$ in the X-direction and $-300^{\prime\prime}$ to $0^{\prime\prime}$ in the Y-direction. In the composite image of the blowout jet, orange colour represents AIA 304 \AA~, green colour reprents AIA 1600 \AA~, and blue colour represents HMI magnetogram.\par
The photospheric magnetic fields at the footpoint of the blowout jet and at its surroundings are shown in the HMI LOS (line of sight) magnetogram at 04:12:38 UT (\textit{cf.} upper panel of Figure 2). The size of this magnetogram is $200^{\prime\prime} \times 200^{\prime\prime}$ with the co-ordinates of $550^{\prime\prime}$ to $750^{\prime\prime}$ in X-direction and $-200^{\prime\prime}$ to $0^{\prime\prime}$ in the Y-direction. We see that there are various quiet Sun magnetic networks in the neighbourhood of the blowout jet, and it is occured from an inter-network quiet Sun element where a circular filament is rooted (\textit{cf.} bottom panels of Figure 3). The footpoint of the blowout jet is at the negative polarity (minority polarity) region in the vicinity of the positive polarity (majority polarity) region (\textit{cf.} upper panel of Figure 2).
For investigating the possible causes of the eruption of the filament and associated blowout jet and to understand the triggering mechanism for their eruptions, we examine the behaviour of the underlying magnetic field. In the initiation phase of the filament (\textit{cf.} Figure 3), we have noticed the initial activity at the northern end of the filament which initiates the eruption of the filament. Therefore, we examine the time evolution of the magnetic flux at around the northern end of the filament.
The time variation of the negative magnetic flux and the positive magnetic flux at the northern end of the filament is shown in the bottom panel of Figure 2. The negative and positive magnetic fluxes are extracted from the box shown by black solid line, which is overplotted on the HMI LOS (line of sight) magnetogram. The size of box is $20^{\prime\prime} \times 15^{\prime\prime}$ with the coordinates of $640^{\prime\prime}$ to $660^{\prime\prime}$ in X-direction and $-135^{\prime\prime}$ to $-120^{\prime\prime}$ in the Y-direction. The magnetic field intensities are extracted for the observational period of 02:59:24 UT - 05:30:54 UT. Negative magnetic flux has the order of $10^{21}$ Mx while positive magnetic flux has the order of $10^{22}$ Mx. The blue dashed line overplotted on this figure indicates the starting time (03:59 UT) of the slow rising phase of the filament. We have noticed that the positive flux shows the declined trend while the negative flux shows an increasing trend. The changing behaviour of the negative and positive magnetic fluxes suggests that the negative flux is emerging and at the same time the flux cancellation between positive and negative flux takes place. This is the confirmation of the magnetic cancellation at the northern end of the filament, where filament eruption and activation of the northern part of the blowout jet are observed. Therefore, we can infer that the magnetic flux cancellation at the northern end of the filament makes it eruptive in multiple parts, which further evolve the blowout jet eruption.
\subsection{ Time-intensity Profile at the Base of the Blowout Jet in Different SDO/AIA Filters}
The lightcurve is plotted in different SDO/AIA filters for analyzing the behaviour of the EUV brightening which is observed at the base of the blowout jet (\textit{cf.} bottom panel of Figure 1). The intensity is extracted in different SDO/AIA filters from the white solid line box which is overplotted on the composite image of SDO/AIA filters over the observational period of 1 hour 40 minute from 03:30:00 UT to 05:10:00 UT. The white-line box size is $60^{\prime\prime} \times 60^{\prime\prime}$ and has the coordinates $630^{\prime\prime}$ to $690^{\prime\prime}$ in the X-direction and $-180^{\prime\prime}$ to $-120^{\prime\prime}$ in the Y-direction. Light curve is plotted in AIA 1600 \AA~ (shown by black colour), AIA 304 \AA~ (shown by red colour), AIA 171 \AA~ (blue colour), AIA 335 \AA~ (yellow colour), AIA 94 \AA~ (violet colour) between the normalized intensity (maximum intensity/mean intensity) and the observational period of the evolution of the blowout jet. Light curve in different SDO/AIA filters show different behaviour i.e., there is no identical intensity peak for all filters. Intensity gets its peak value in AIA 1600 \AA~, AIA 304 \AA~ earlier in comparison to other filters. For these two filters intensity peaks show nearly same behaviour as first peak observed at about 04:08:18 UT and second peak is observed at about 04:11:36 UT. This demonstrates that cool plasma is firstly evolved during the formation of the blowout jet's spire. This is an opposite scenario as in typical coronal jets the hot plasma evolves first and the cool plasma thereafter (e.g., Jiang et al. 2007, Nishizuka et al. 2008, Solanki et al. 2018). This current observations reveal that some distinct mechanism is at work in the formation of this quiet-Sun blowout jet other than the typical magnetic reconnection in the corona. The most likely scenario is a collective small-scale flux emergences and subsequent cancellation with the neighbourhood at the boundary of magnetic network (\textit{cf.} Figure 2). This launches the bulk plasma flows into the pre-existing blowout jet spire's magnetic field in the upward direction.
The cool plasma consists of the temperature range from $\approx 10^{4}$ - $10^{5}$ K. After these two filters intensity get its peak value in AIA 94 \AA~ and in AIA 335 \AA~ at about 04:20:00 UT and 04:23:18 UT respectively. AIA 171 \AA~ shows slightly different kind of behaviour in intensity plot as there are many peaks in light curve of AIA 171 \AA~, first two small peaks match well with AIA 1600 \AA~ and AIA 304 \AA~ at 04:08:18 UT and at 04:11:36 UT with less intensity. Intensity gets its peak value at about 04:33:18 UT in AIA 171 \AA~. In the light curve, the shift is observed in the intensity peaks of different SDO/AIA filters, which emphasizes the time-lagging behaviour of the evolution of multi-temperature plasma throughout the entire period of the evolution of blowout jet at different time epoch . This time-lagging behaviour indicates the presence of flare evolution at the base of blowout jet. A weak flare is observed at about 04:08 UT near to the northern side of the blowout jet and the filament, which accelerates the plasma to the eruption. We relate this weak flare with the network flare (Krucker et al 1997, Krucker et al. 2000). \par
\subsection{Evolution of the Blowout Jet due to the Eruption of Segments of a Filament as seen in Different SDO/AIA Filters and GONG H$\alpha$}
\par
The initiation phase of the blowout jet in different SDO/AIA filters as seen in AIA 304 \AA~, AIA 171 \AA~ and in GONG H$\alpha$ 6563 \AA~ indicates the presence of the cool plasma and the filament at the base of the blowout jet (\textit{cf.} bottom panels of Figure 3). This filament is embedded at the internetwork region, which can be seen in first image of bottom panel in H$\alpha$ at 04:03:54 UT. In the GONG H$\alpha$ observations this filament looks like circular shaped structure. Firstly a slow rise is observed in the filament. The cool plasma and ustable filament move up and the jet bright points are created at the filament root at 04:08:54 UT, which is a network flare (see Figure 3). The filament is erupted at about 04:10:54 UT. In the initiation phase, filament shows slow rise, ejection and evolution of the blowout jet. The evolution of the blowout jet can be seen in the animation Movie1.\par
The filament is ejected in two stages as the circular shaped structure of the filament divides into two parts. In the first stage, the eruption of northern part of the filament (first part) takes place and drives the blowout jet (\textit{cf.} Figure 4). The hot plasma escapes out and moves linearly along the open magnetic field lines and form a broad and complex jet-like spire (\textit{cf.} northern part of eruptions in Figure 4). In the blowout jet, eruption of the northern side of the blowout jet is most activated in it's primary phase, and significant plasma dynamics is seen along it (\textit{cf.} Figure 4). The significance of the formation of plasma blobs are clearly evident in this part, which we will describe in the forthcoming sub-section 3.4.\par
\subsection{Formation of the Plasma Blobs in the Northern Spire of the Blowout Jet}
We have found the plasma blobs at the edge of the northern spire of the blowout jet. The formation and evolution of these plasma blobs can be seen in the time-sequence images of AIA 304\AA~ in Figure 5, 6. The selected FOV for these images is $80^{\prime\prime} \times 50^{\prime\prime}$ from $670^{\prime\prime}$ to $750^{\prime\prime}$ in X-direction and from $-150^{\prime\prime}$ to $-100^{\prime\prime}$ in the Y-direction in Figure 5. These plasma blobs are identified as B1, B2, B3, B4 in the middle panel of Figure 5. The total time-duration of these plasma blobs are about 03 min 48 sec. The northern side of the blowout jet where these plasma blobs are formed is highlighted with the blue-lined box in Figure 6. The bottom panel of Figure 6 shows the zoomed picture of the plasma blobs. The formation of these plasma blobs results from the Kelvin-Helmholtz (K-H) instability. These hot and high-dense plasma blobs are moving along the northern spire of the blowout jet. The K-H instability arises due to the shear flows between the high speed blowout jet and the steady local plasma and results into the formation of the magnetic islands. These islands take the form of plasma blobs at the time of the evolution. \par
We have analyzed the velocity field at the northern side of the blowout jet by using Fourier ocal Correlation Tracking (FLCT; Fisher et al. 2008) method (\textit{cf.} right panel of Figure 7). The velocity field is analyzed at the northern side of the blowout jet corresponding to the box which is overplotted on the AIA 1600 \AA~ image having the coordinates of $500^{\prime\prime}$ to $900^{\prime\prime}$ in the X-direction and $-300^{\prime\prime}$ to $0^{\prime\prime}$ in the Y-direction (\textit{cf.} left panel of Figure 7). We have selected two HMI LOS magnetograms first magnetogram at 03:37:36 UT and second magnetogram at 04:14:08 UT (at the timing of the onset of the blowout jet) to estimate the flow field velocity field. In the right panel of Figure 7, we see the base image of HMI magnetogram at 03:37:36 UT and the velocity field which is shown by orange arrows overplotted on the base image as taken from HMI magnetogram at 04:14:08 UT. \par
It is clear that the clock-wise plasma flows is evident centered at that particular quiet-Sun region from where the blowout jet is originated.
The right panel of Figure 7 shows the partial field of view showing the footpoint of the northern side of the blowout jet (\textit{X} = $640^{\prime\prime}$ - $710^{\prime\prime}$,\textit{Y} = $-125^{\prime\prime}$ - $-100^{\prime\prime}$). Yellow contour which is overlaid on HMI magnetogram shows the AIA 304 \AA~ intensities. This is clear that the clock-wise shearing flow-field is acting at the footpoint of the northern side of the blowout jet, which further launch the right-handed twist in the entire overlying plasma column/spire associated with the blowout jet. When this clock-wise shearing flow-field moves further and interact with the local stationary plasma field, it causes the K-H instability which results in the formation of four plasma blobs as B1, B2, B3, B4. The plasma blobs are less visible in AIA 304 \AA~ channel due to the less spatial and time resolutions data. The northern spire of the jet also does untwisting/rotational motion and releases its twist. The jet's spire shows the estimated twist about 1-1.5 turns (or $2\pi$-$3\pi$). Our twist value is found in good agreement with the results of the Pariat et al. 2009, which shows for the driving of a solar coronal jet the threshold value of the twist is 1.4 turns ($2.8\pi$). Our finding of the K-H unstable plasma blobs in the rotating/untwisting spire of the blowout jet supports the numerical results of the Ni et al. 2017, Zhelyazkov et al. 2018 and Zhelyazkov et al. 2018. This K-H unstable northern plasma spire further moves into higher coronal region and drives a jet-like CME, which we will discuss in the forthcoming Section 3.7.\par
\subsection{Kinematics of the Northern Part of the Blowout Jet}
We have done the height-time analysis of the northern part of the blowout jet. To calculate the height of the blowout jet, we have used the tie-pointing method of the Inhester et al. 2006. In this method the triangulation technique between the different view points of the STEREO\_A and SDO/AIA is used. In this triangulation technique, scc\_measure.pro is used which is available in solarsoft library for estimation of real height of blowout jet. 
We have calculated the height of the blowout jet by tracking the tip of the blowout jet in two different view points of the STEREO\_A and SDO/AIA (\textit{cf.} upper panel of Figure 8). In the bottom panel of Figure 8, the height of the jet has been tracked using its tip starting at 04:06:09 UT when the jet is clearly visible in both the instruments (SDO/AIA 304 \AA~ and STEREO\_A EUVI 304 \AA~) while the initiation of jet takes place at 03:59 UT. We have calculated the height (in $R_\odot$) of the blowout jet at ten different times (in UT) as 04:06:19, 04:10:31, 04:14:07, 04:15:31, 04:16:19, 04:16:31, 04:20:31, 04:25:31, 04:26:15, 04:30:30.  With this height-time array, we have calculated the velocity of the blowout jet which is $325$ $km sec^{-1}$ (\textit{cf.} bottom panel of Figure 8). The calculated acceleration for this blowout jet is $-0.30$ $km sec^{-2}$ which is estimated by the second order fitting in the H-T plot of blowout jet.
\subsection{Eruption of the Southern Part of the Filament and Associated Blowout Jet}
After the activation and eruption of the northern section of the filament and associated segment of the blowout jet, the southern section of the filament is also erupted in the form of a twisted/deformed flux rope (\textit{cf.} H$\alpha$, AIA 304 \AA~ images in bottom and middle panel of Figure 9 which makes the second stage of the eruption (\textit{cf.} AIA 171\AA~ images of Figure 9).  The eruption of the southern part of the filament can be seen in the animation Movie1. This deformed/twisted magnetic flux rope moves further and releases its helicity and forms a rotating plasma spire of the southern part of the blowout jet.
The eruption of the northern section of the filament destabilizes and removes the local magnetic field configurations and induces the eruption of the southern section of the filament. Alternatively we may adopt the magnetic implosion physical mechanism for the initiation of the eruption of the southern section of the filament. It may occur between neighbouring magnetic structures confined by a large magnetic structure (e.g., Hudson 2000, Liu et al. 2009, Shen et al. 2012). At the magnetic implosion site, the eruption takes place when the upward magnetic pressure decreases resulting into the contraction of overlying field and free magnetic energy release.
\subsection{Jet-like and Bubble-like Twin CME}
In this analyzed event, we have found the generation of twin CME associated with the blowout jet eruption. Northern part of the filament erupts firstly and causes the evolution of the coronal blowout jet, which may also subject to the K-H instability. Thereafter, the southern part of the filament also erupts in form of magnetic flux rope, and form the full blowout jet eruption. \par
The northern and southern segments of the filament may be confined by the same magnetic field system. The magnetic flux cancellation is occured at the northern end of the filament which makes this filament unstable and erupts it in different stages. In first stage northern segment of filament ejects and initiates the eruption of the northern part of the blowout jet. The northern part of the blowout jet further drags the first CME which is a jet-like CME. This jet-like CME is the extension of the collimated plasma beam which is generated by the external magnetic reconnection (Shen et al. 2012).\par
The eruption of the northern section of the filament removes the confined magnetic field of the southern section of the filament and induces the eruption of the southern section of the filament. Alternatively we may adopt the magnetic implosion mechanism for the eruption of the southern section of the filament. The eruption of the southern section of the filament form the full blowout jet eruption and causes the second CME which is a bubble-like CME. \par
The line of sight (LOS) evolution of these twin CME are shown in running difference images of SoHO/LASCO C2 coronagraph (\textit{cf.} Figure 10). The dynamics of these two CMEs can be seen in the animation Movie2. We have used the multi-scale gaussian normalization method of Morgan et al. 2014 in making these running difference images. We have marked the jet-like CME and bubble-like CME in Figure 10 in the image of 05:12:05 UT. The bubble-like CME has typical three part structures as bright core which consists of cool plasma material of the filament, dark cavity and the bright front of the CME. The bright core is at the northern side of the dark cavity, generally the bright core resides at the central of the dark cavity and the CME. The first appearance of the CMEs in the LASCO C2's FOV is observed at about 04:38:53 UT. The spatial and the temporal relationship between the twin CME and the northern as well as southern part of the jet eruption relates to the eruptions of the northern and southern sections of the filament which determine jet-like and the bubble-like CME.
\subsection{Kinematics of the Jet-like and Bubble-like CME}
We have done the height-time analysis for the study of the kinematics of jet-like and bubble-like CMEs. We have applied the same method for height-time measurements of these twin CME. We have measured the projected height for these CMEs with respect to the centre of the solar disk by using the tie-pointing method of the Inhester et al. 2006. We have tracked the tip of the CMEs in STEREO\_A COR2 and STEREO\_B COR2 simultaneously with the help of the triangulation technique. The separation angle between STEREO\_A and STEREO\_B on 2014 May 16 at 04:00 UT is about \textdegree{36}. Since longitude angle is high in measurements, we have estimated the projected height of the twin CME. In case of jet-like CME, we have estimated the projected height at five different times as 05:09:15 UT, 05:24 UT, 05:39 UT, 05:54 UT and 06:09:15 UT. At each time we have tracked the tip of the CME (measurement of projected height) at ten times to estimate the underlying uncertainty in the measurements. With these data sets of projected height and time we have plotted the H-T plot for jet-like CME (\textit{cf.} Figure 11). The calculated velocity for jet-like CME is about $619$ $km sec^{-1}$. We have done the second order fitting on the H-T plot to calculate the acceleration of the CME and get the $0.35$ $km sec^{-2}$ acceleration value for the jet-like CME. \par
In case of bubble-like CME we have estimated the projected height at 04:54 UT, 05:09:15 UT, 05:24 UT, 05:39 UT, 05:54 UT, 06:09:15 UT, 06:24 UT, 07:09:15 UT. With the projected height and time data sets we have plotted the H-T plot for bubble-like CME and calculated the velocity of the bubble-like CME (\textit{cf.} Figure 12). The calculated velocity is about $620$ $km sec^{-1}$.
The calculated acceleration is about $-0.031$ $km sec^{-2}$. We have done the second order fitting on the H-T plot of bubble-like CME to get the value of this negative acceleration.
\section{Discussion and Conclusions}
There are many observational results which deal with the blow-out jet eruption from the active region of the Sun and their different characteristics and triggering mechanisms.
Li et al. 2015 observed an active-region blow-out jet which is associated with a CME and a M-class solar flare , where the filament eruption triggers this blow-out jet. Li et al. 2017 observed a blow-out surge in coronal loops where he found that the magnetic reconnection between the erupting filament and the coronal loop is responsible cause for the eruption of blowout surge. Shen et al. 2017 observed an active region blow-out jet with SDO, this jet is associated with the filament eruption and consisted of hot and cool plasma structure, where cool plasma component preceeds further than the hot plasma component. Hong et al. 2017 studied an active region blow-out jet associated with C-class flare and a type-\textsc{iii} radio burst, where it is observed that the filament eruption triggers these eruptive events. Li et al. 2018 has recently discovered the Kelvin-Helmholtz Instability (KHI) in an active region penumbral structural blow-out jet in the high-resolution Interface Region Imaging Spectrograph (IRIS) observations. \par
In the present work, we describe the physical properties, triggering mechanism and kinematics of a quiet-Sun blowout jet which is observed by SDO/AIA in different wavelengths on 2014 May, 16. This blowout jet is initiated by the multi-section eruptions of the circular filament which is at the base of the blowout jet. Based on our observational results, we have also found the generation of the twin CME as jet-like and bubble-like CMEs which are associated with blowout jet eruption.\par
Here are some concluding points of this observed event in the studied observed baseline. \par
i) In the time-intensity profile of the blowout jet it is observed that the cool plasma has the temperature range $\approx 10^{4}$-$10^{5}$ K, which is evolved earlier in the formation of the magnetized plasma spire of the blowout jet. This is due to the multiple filament ejection which further drags blowout jet. This is the unique scenario which shed the light that the formation mechanism of this blowout jet is entirely different from the typical coronal reconnection driven blowout jets.
There is the presence of the multi-temperature plasma and the time-lagging behaviour is observed during the evolution period of the blowout jet, which is analogous with the flare eruption. We relate this flare with the network flare (Krucker et al. 1997, Krucker et al. 2000). This flare energy accelerates the blowout jet plasma. The energisation of this network flare is due to the activation of filament segments and its reconnection with the existing overlying fields.\par
ii) The time-evolution of the magnetic flux at the northern end of the filament is analyzed which shows the magnetic cancellation signature. The magnetic cancellation destabilizes the filament and further makes it to erupt in different stages.  \par
iii) The complete evolution of the blowout jet and the filament is observed in AIA 304 \AA~ (at TR temperature), AIA 171 \AA~ (at inner coronal temperature) and in H$\alpha$ (at chromospheric temperature). This eruption goes through the different stages as the circular filament ejects in two stages. Firstly the northern section of the filament lifts up, ejects and drives the northern part of blowout jet. In second stage the southern section of the filament also erupts and forms the rotating plasma spire of the blowout jet, \textit{i.e.} southward part of the blowout jet. \par
iv) The plasma blobs are formed at the edge of the northern plasma spire of the blowout jet which are moving along the jet's spire. These plasma blobs are most likely subjected to the K-H instability, which arises due to the interaction between the sheared motion of the northern part of the blowout jet and the local stationary plasma in the surrounding. \par
v) The velocity field is analyzed at the footpoint of the northern part of the blowout jet using Fourier Local Correlation Technique (FLCT). The velocity field shows the clock-wise plasma flows is centered at the blowout jet triggering site, which enables the magnetic twists of similar sign in the whole northern spire of the blowout jet. This enables the sheared plasma motion in the jet's spire and most likely the evolution of K-H unstable plasma blobs. \par
vi) We have done the height-time analysis of the northern side of the blowout jet. The calculated velocity and acceleartion are found to be $325$ $km sec^{-1}$ and $-0.30$ $km sec^{-2}$ respectively. \par
vii) The twin CME are observed associated with the blowout jet. The eruption of the northern part of the blowout jet drives the jet-like CME. The outward moving hot plasma on the disk is observed as northern part of the blowout jet and in the outer coronal region it is observed as the jet-like CME. The jet-like CME is the extension of the collimated plasma beam which is generated by the external magnetic reconnection (Shen et al. 2012). The eruption of the southern section of the filament enables the rotating spire of the blowout jet, which further drives the bubble-like CME. These twin CME are observed simultaneously. \par
viii) The calculated velocity and acceleration for jet-like and bubble-like CMEs are found to be $619$ $km sec^{-1}$ and $0.35$ $km sec^{-2}$, $620$ $km sec^{-1}$ and $-0.031$ $km sec^{-2}$. respectively\par
In the observations of the Shen et al. 2012 the double CMEs are less distinguishable but in our case, we can easily distinguish the jet-like and bubble-like CME as these CMEs occur side by side. To the best of our knowledge, our observed event is the third event of the twin CME with blowout jet eruption after the observations of the Shen et al. 2012 and Miao et al. 2018. 
\begin{figure*} 
 \centerline{\includegraphics[width=10cm]{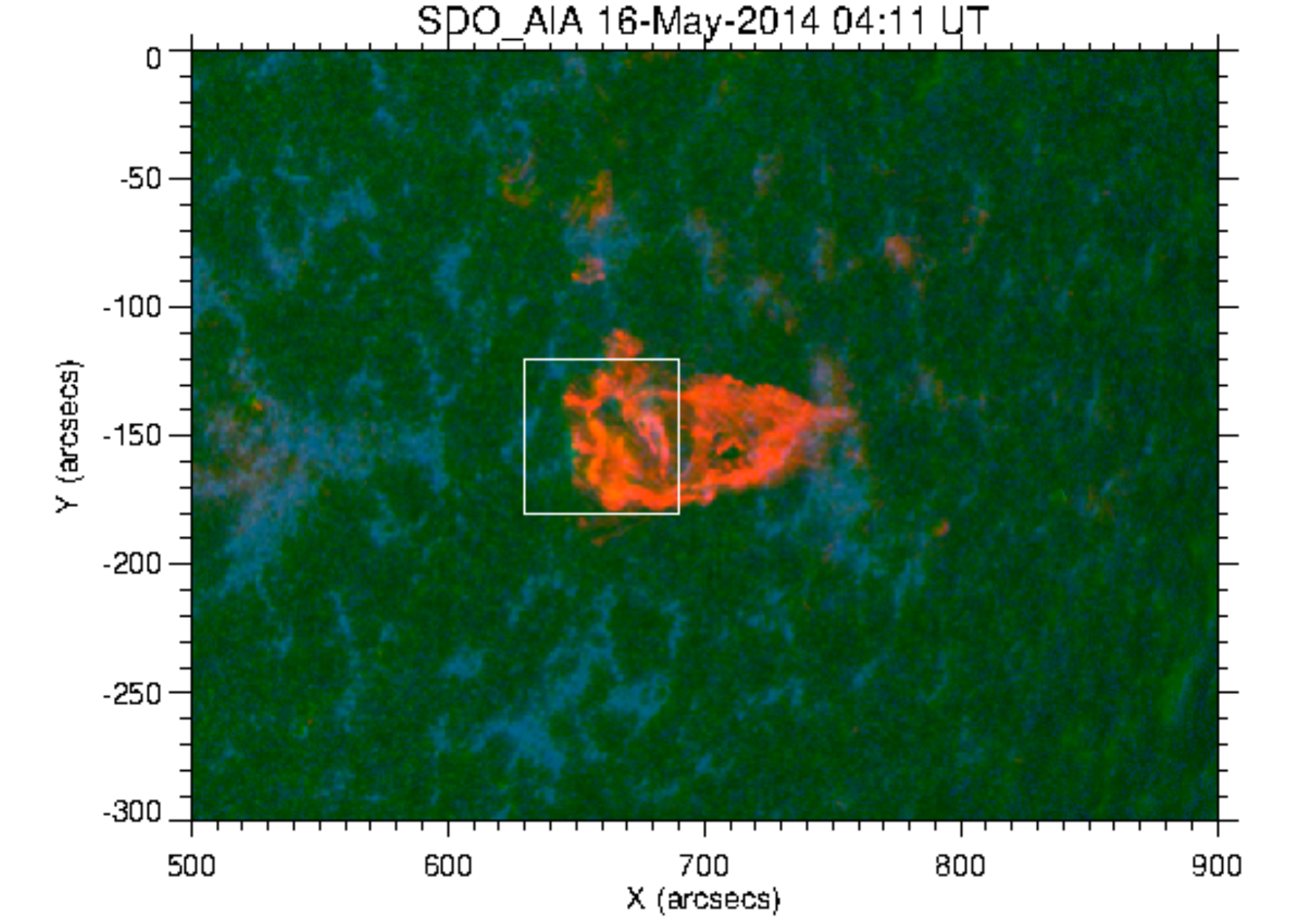}}
             \vspace*{-0.1\columnwidth}
\end{figure*}
\begin{figure*}
 \centerline{\includegraphics[width=8cm,angle=90]{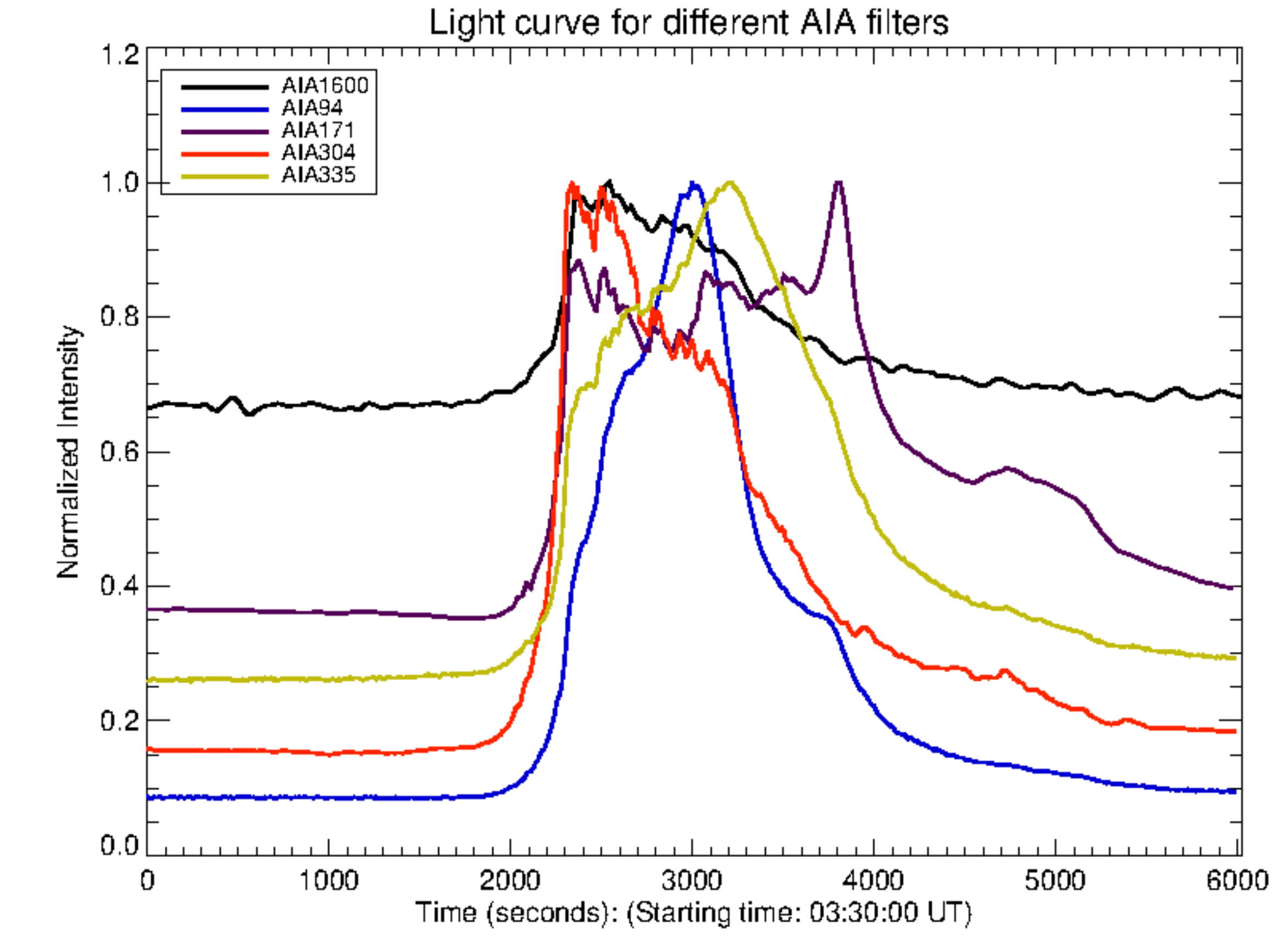}
}
\caption{Upper panel shows the blowout jet eruption in the composite image of SDO/HMI, AIA 1600 \AA~, and AIA 304 \AA~ at 04:11:00 UT. Bottom panel shows light curves obtained from different filters of SDO/AIA at the footpoint of the blowout jet.}
\label{Figure 1}
\end{figure*}
\begin{figure*}
\centerline{\includegraphics[height=8cm]{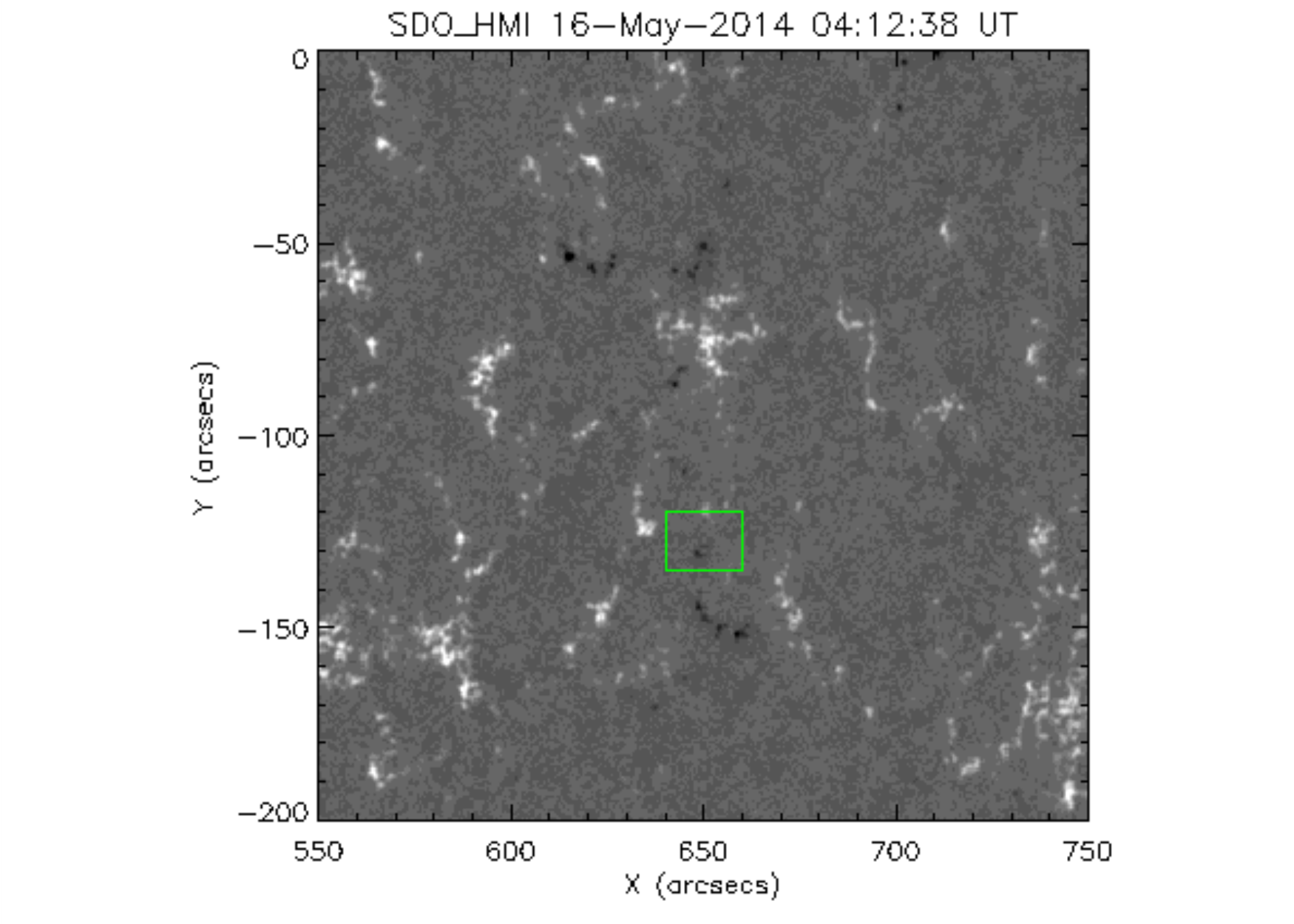}}
\vspace*{-0.1\textwidth}
\end{figure*}
\begin{figure*}
\centerline{\includegraphics[angle=90,height=8cm]{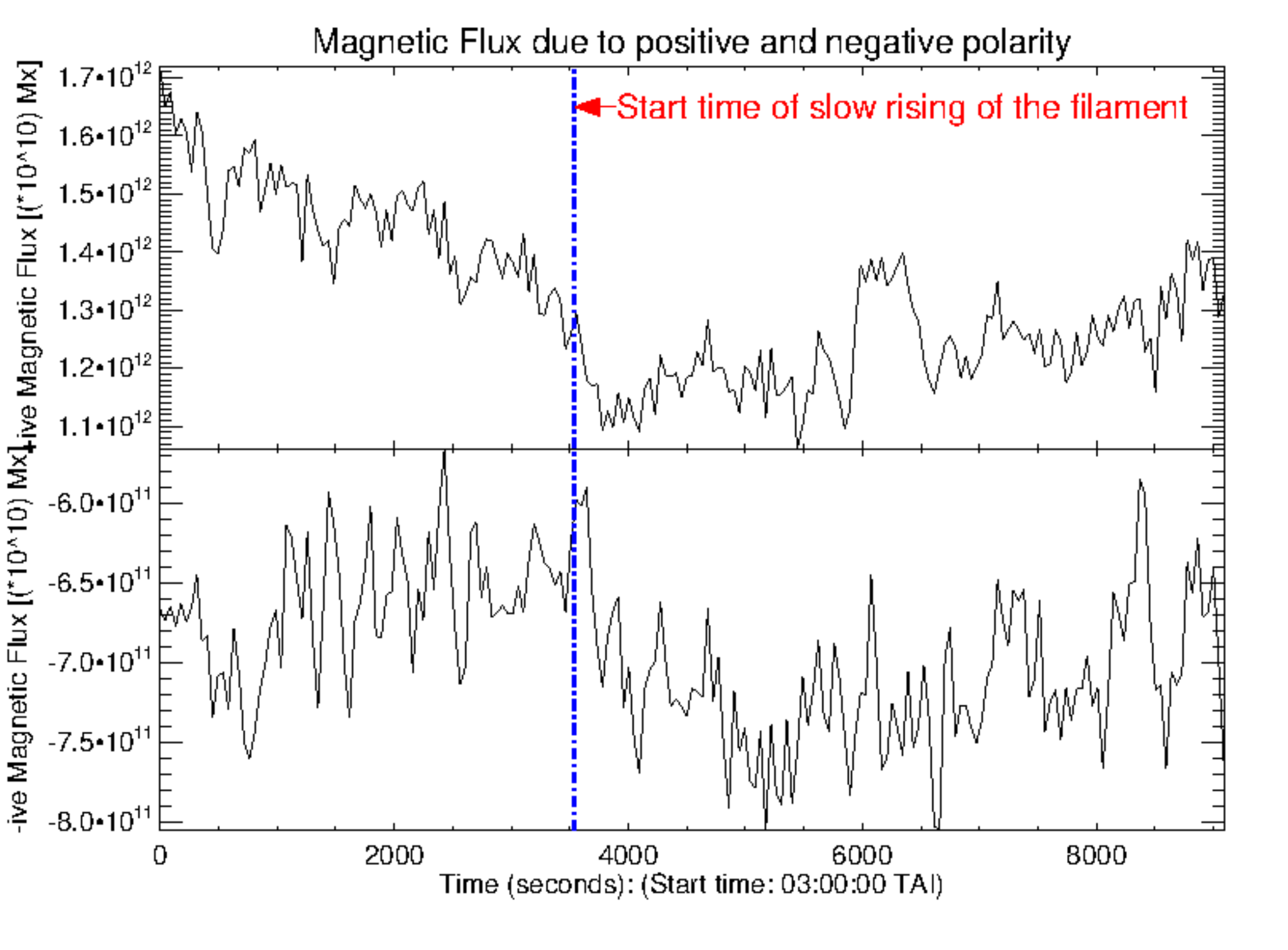}}
\caption{Upper panel shows SDO/HMI line-of-sight (LOS) magnetogram at 04:12:38 UT. Bottom panel shows the temporal evolution of positive and negative magnetic flux corresponding to the overplotted box on HMI magnetogram of size  $20^{\prime\prime} \times 15^{\prime\prime}$ at around the northern end of the filament.}
\label{Figure 2}
\end{figure*}
\begin{figure*}
\centerline{\includegraphics[height=15cm,width=18cm]{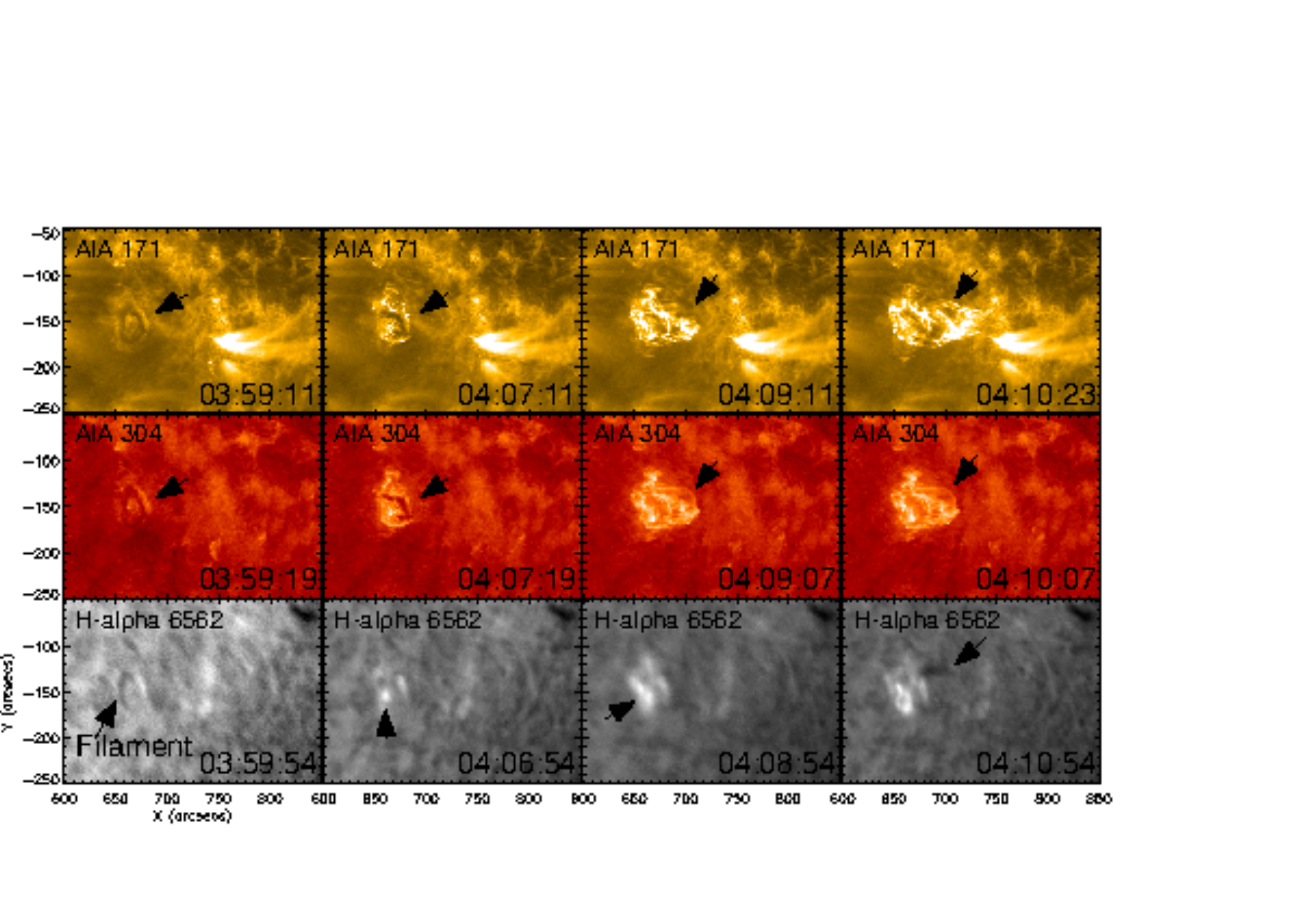}}
\caption{The complete picture of the initiation of the initial phase of the blowout jet due to the activation of circular filament. Northern section of the filament starts to eject at 04:10:54 UT in H$\alpha$ and evolves the eruption of the blowout jet (\textit{cf.} Movie1; top-most row).}
\label{Figure 3}
\end{figure*}
\begin{figure*}
\centerline{\includegraphics[height=15cm,width=18cm]{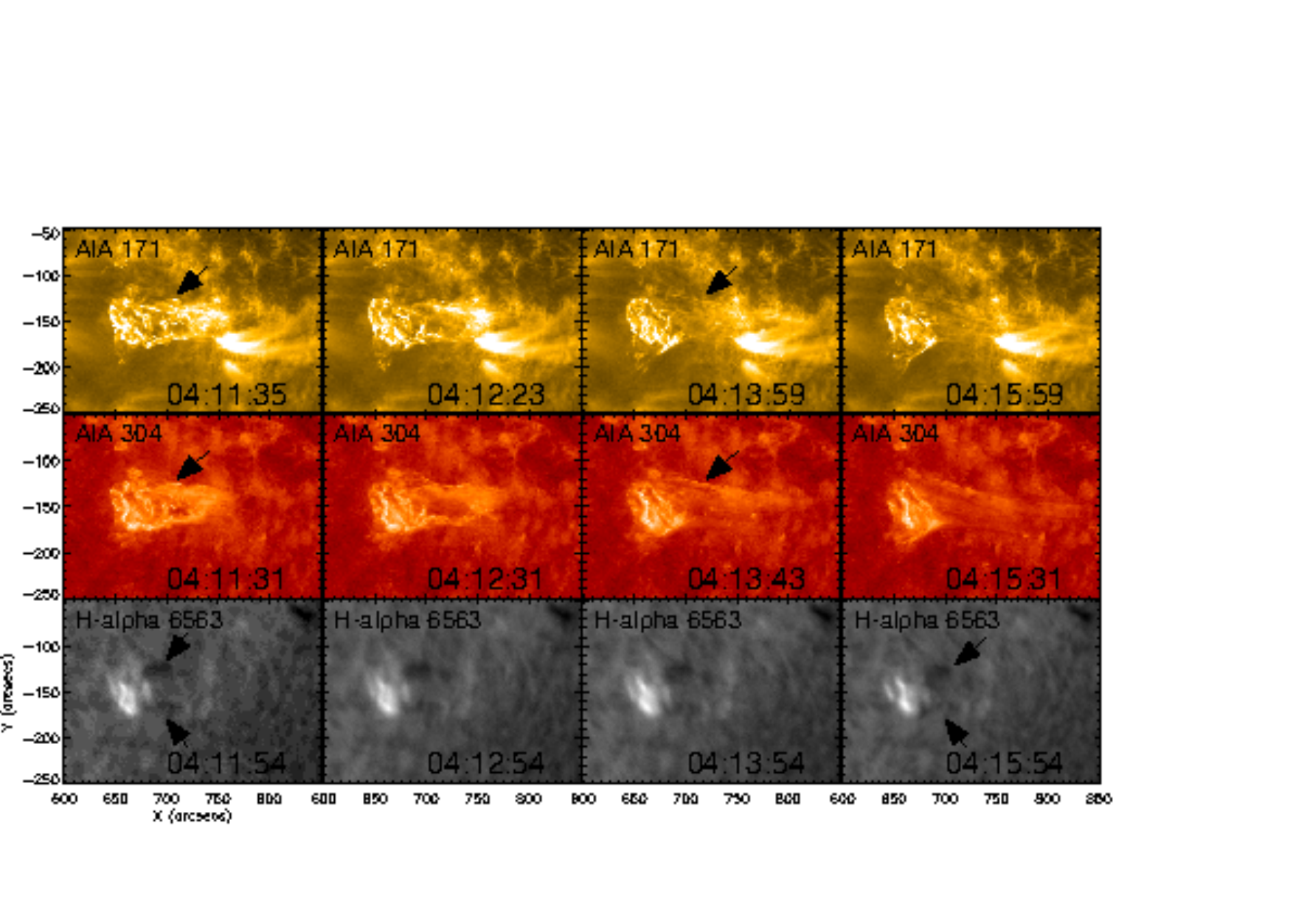}}
\caption{Evolution of northern part of the blowout jet in the time-sequence images of the AIA 304\AA~, AIA171 \AA~, and H$\alpha$. The hot plasma escapes and form a broad, complex, northern spire of blowout jet which does linear motion, and exhibits the formation of plasma blobs (\textit{cf.} Movie1; middle row).}
\label{Figure 4}
\end{figure*}
\begin{figure*}
\centerline{\includegraphics[width=15cm]{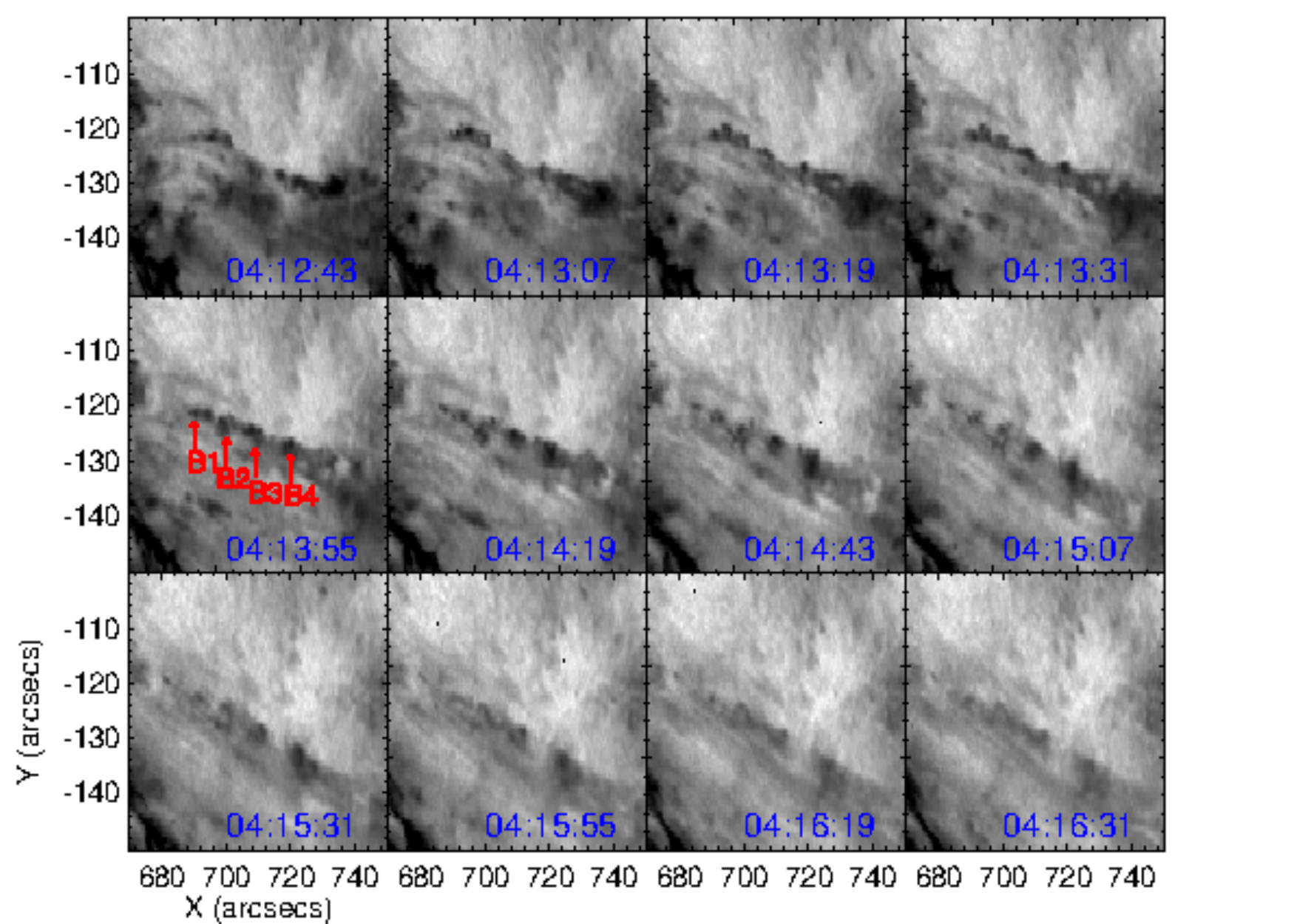}}
\caption{The plasma blob formation in the northern side of the blowout jet is shown in the time-sequence images of SDO/AIA 304\AA~.}
\label{Figure 5}
\end{figure*}
\begin{figure*}
\centerline{\includegraphics[height=11cm,width=9cm,keepaspectratio]{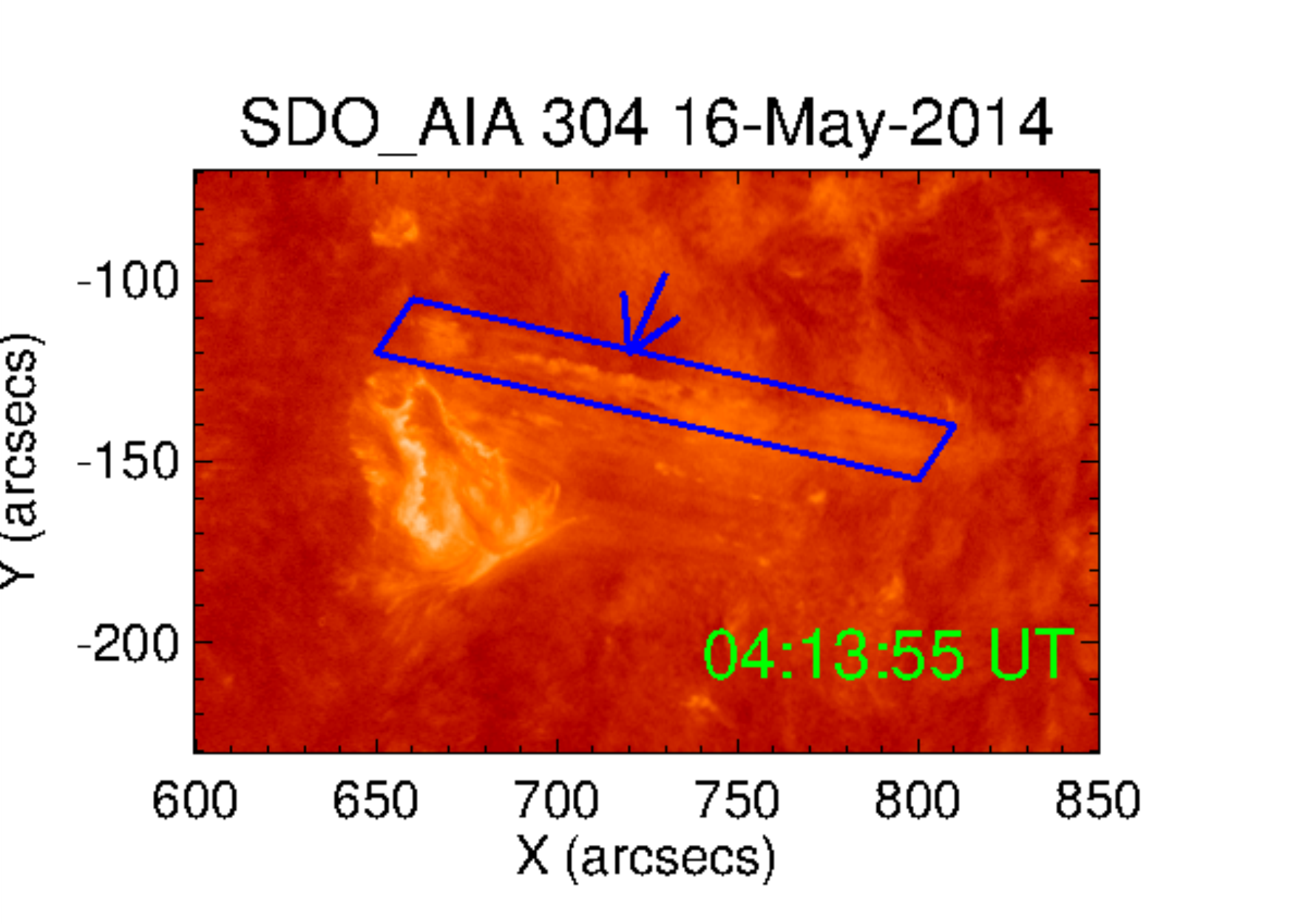}}
\vspace*{-0.03\textwidth}
\centerline{\includegraphics[height=16cm]{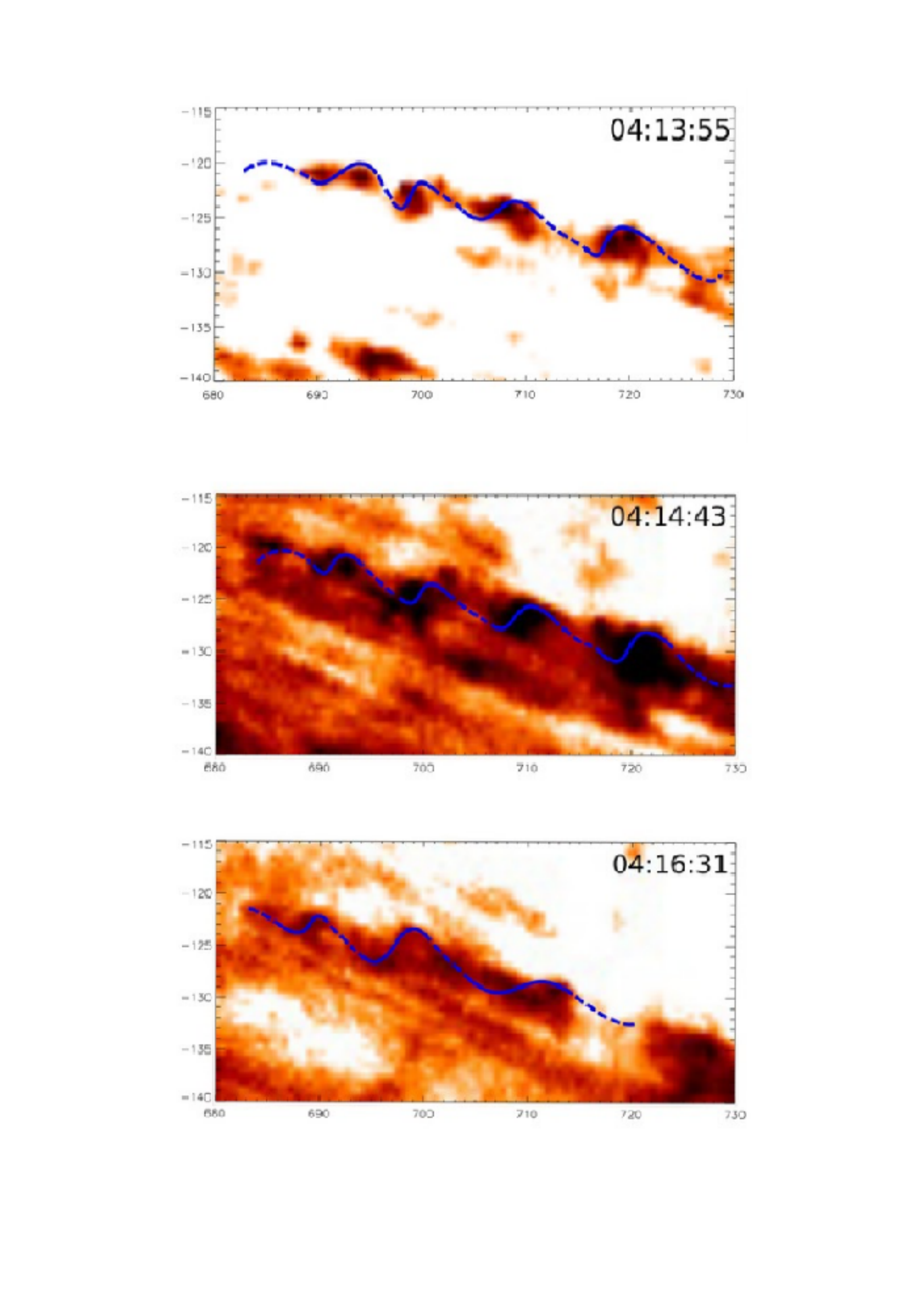}}
\caption{In the upper panel blue-line box highlighted the northern side of the blowout jet, where formation of plasma blobs takes place. In the bottom panel images of SDO/AIA 304 \AA~ show the evolution of plasma blobs as well as launch of the magnetic twists at different epoch.}
\label{Figure 6}
\end{figure*}
\begin{figure*} 
\vspace{0.1\textwidth}
\centerline{\includegraphics[height=13cm,width=13cm,keepaspectratio]{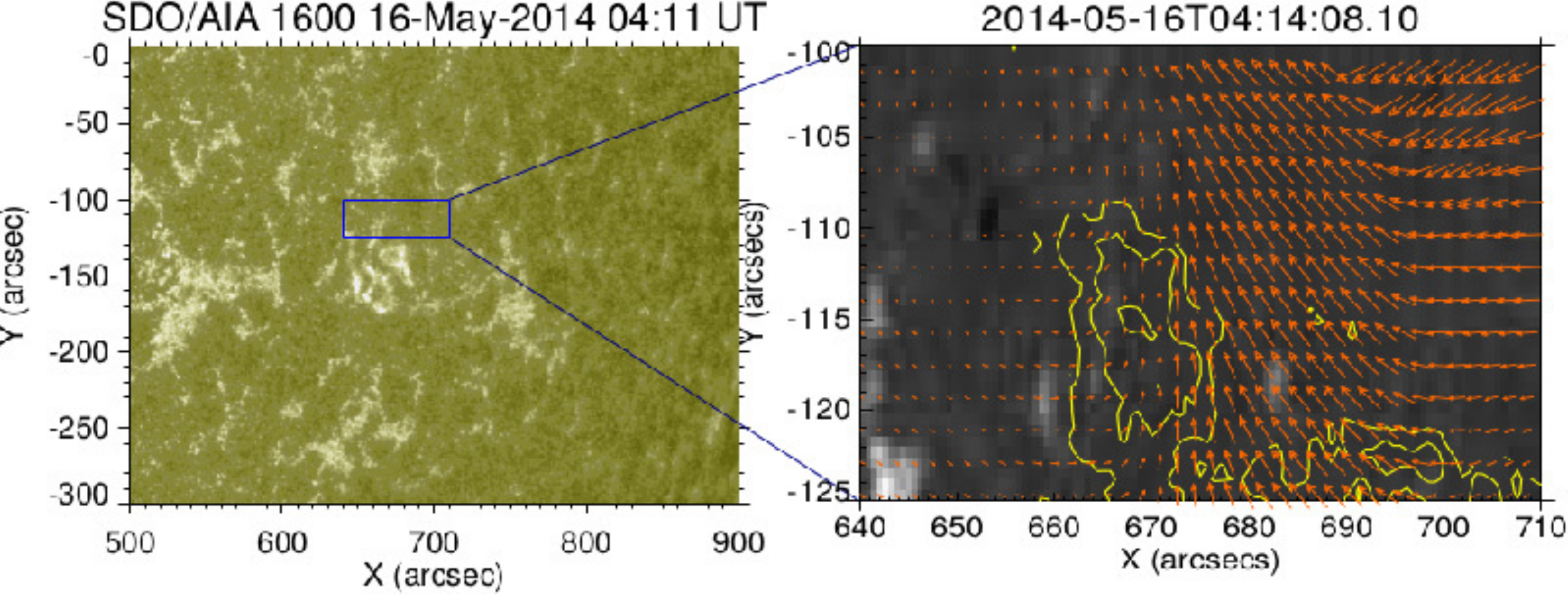}}
\caption{Left panel shows AIA 1600 \AA~ image at 04:11 UT, a blue solid line box is overplotted on it which indicates the northern side of the blowout jet. Right panel shows the velocity field at the northern side of the blowout jet, where clockwise plasma shearing motion is evident.}
\label{Figure 7}
\end{figure*}
\begin{figure*}
\centerline{\includegraphics[height=14cm,width=12cm,keepaspectratio]{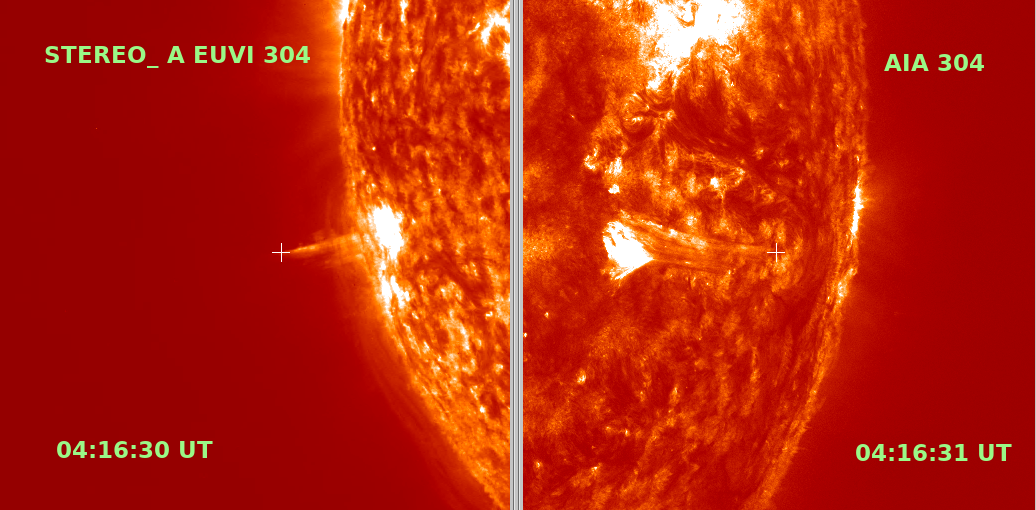}}
\vspace*{0.03\textwidth}
\centerline{\includegraphics[height=12cm,width=11cm,keepaspectratio]{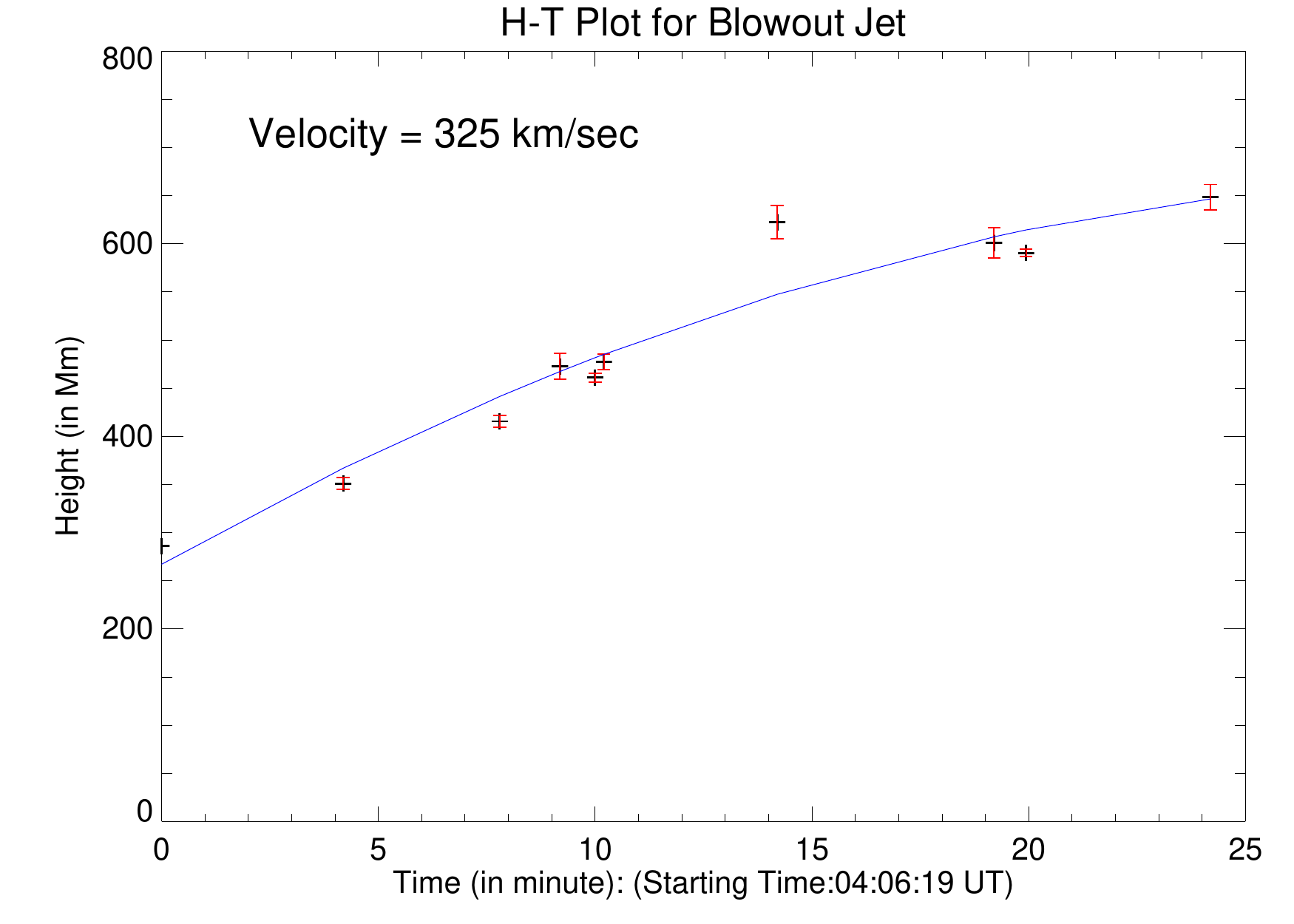}}
\caption{In upper panel tip of the blowout jet is tracked simultaneously in STEREO\_A EUVI 304 \AA~ and AIA 304 \AA~ by using the triangulation technique. In bottom panel the height-time plot for blowout jet is shown.}
\label{Figure 8}
\end{figure*}
\begin{figure*}
\centerline{\includegraphics[height=15cm, width=18cm]{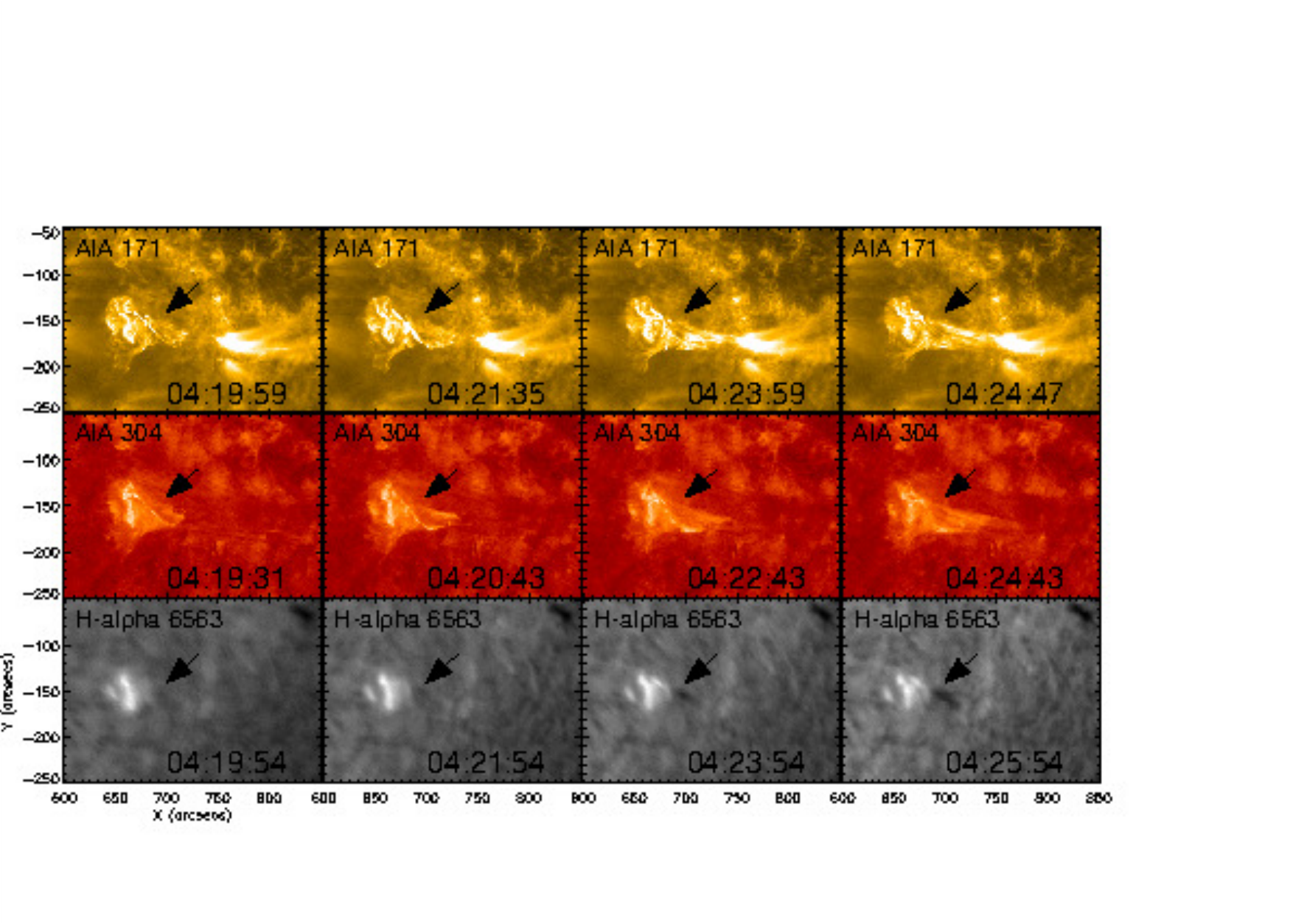}}
\caption{In second stage, the eruption of southern section of the filament is occured in form of twisted/deformed magnetic flux rope. This enables the formation of rotating plasma spire of the southern segment of this blowout jet (\textit{cf.} Movie1; bottom row).}
\label{Figure 9}
\end{figure*}
\begin{figure*}
\centerline{\includegraphics[angle=90,height=14cm,width=17cm]{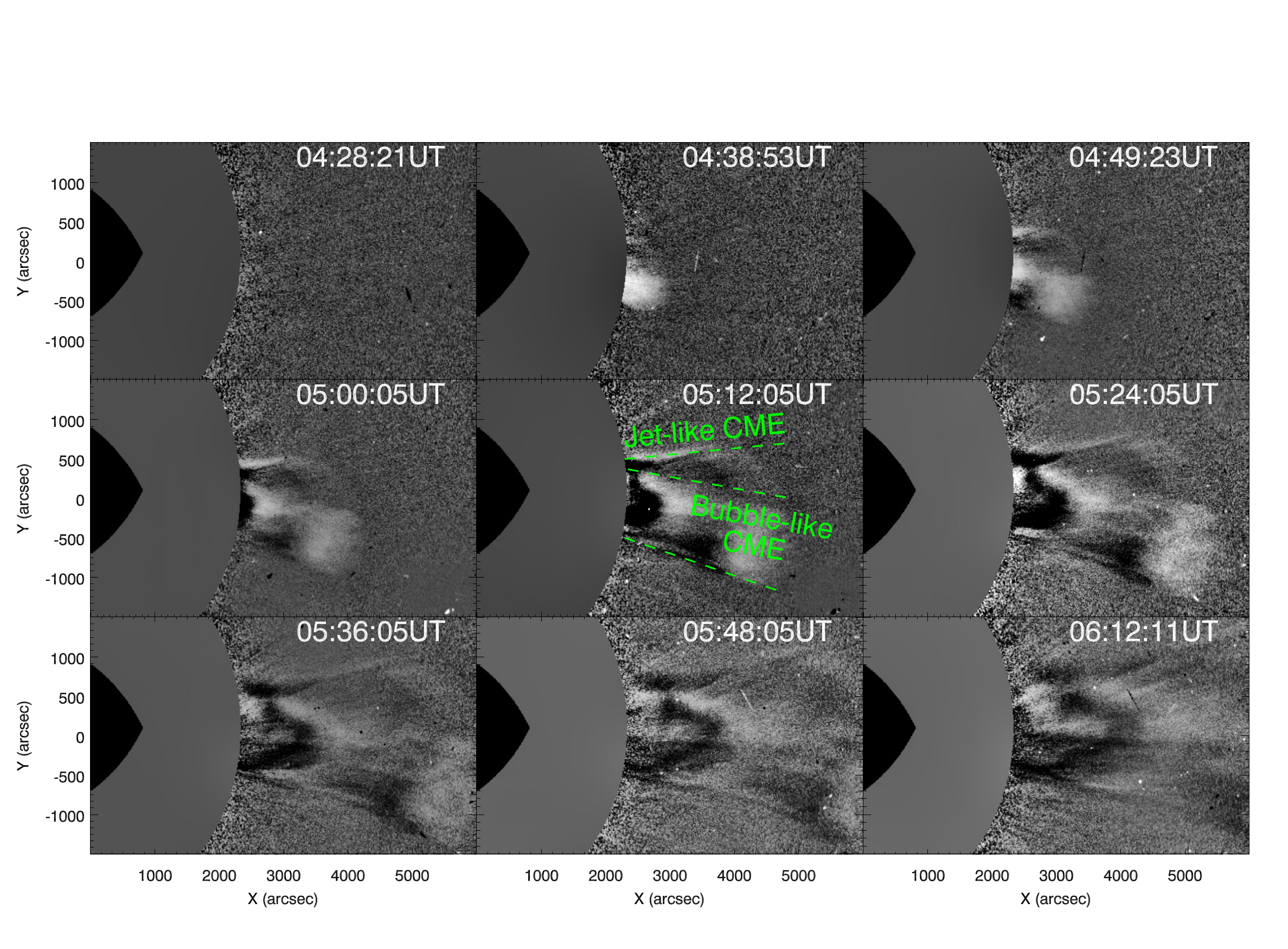}}
\caption{Twin CME evolution at different times as observed in SoHO/LASCO C2 coronagraph (\textit{cf.} Movie2). The northern part of the blowout jet drages the jet-like CME and the eruption of the southern section of the filament causes the bubble-like CME.}
\label{Figure 10}
\end{figure*}
\begin{figure*}
\centerline{\includegraphics[height=14cm,width=12cm,keepaspectratio]{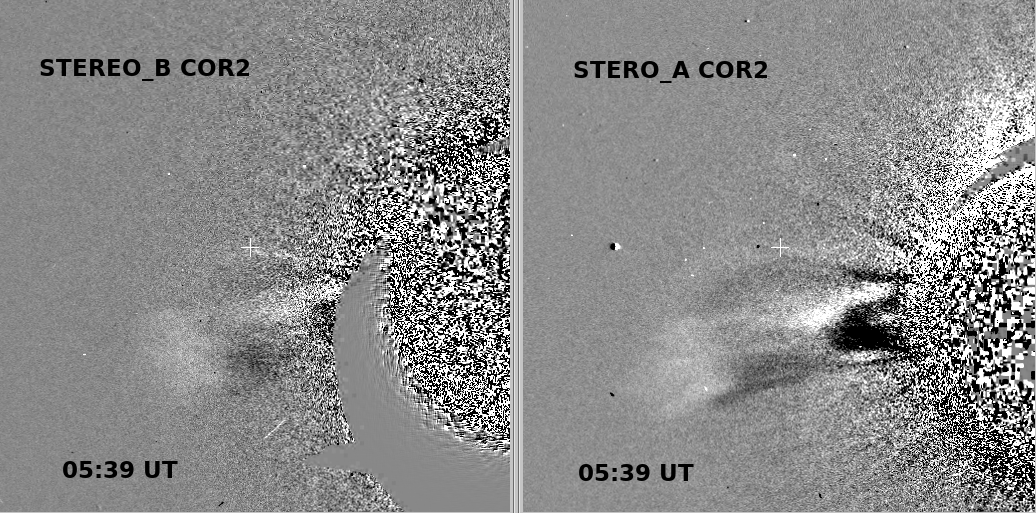}}
\vspace*{0.055\textwidth}
\centerline{\includegraphics[height=12cm,width=11cm,keepaspectratio]{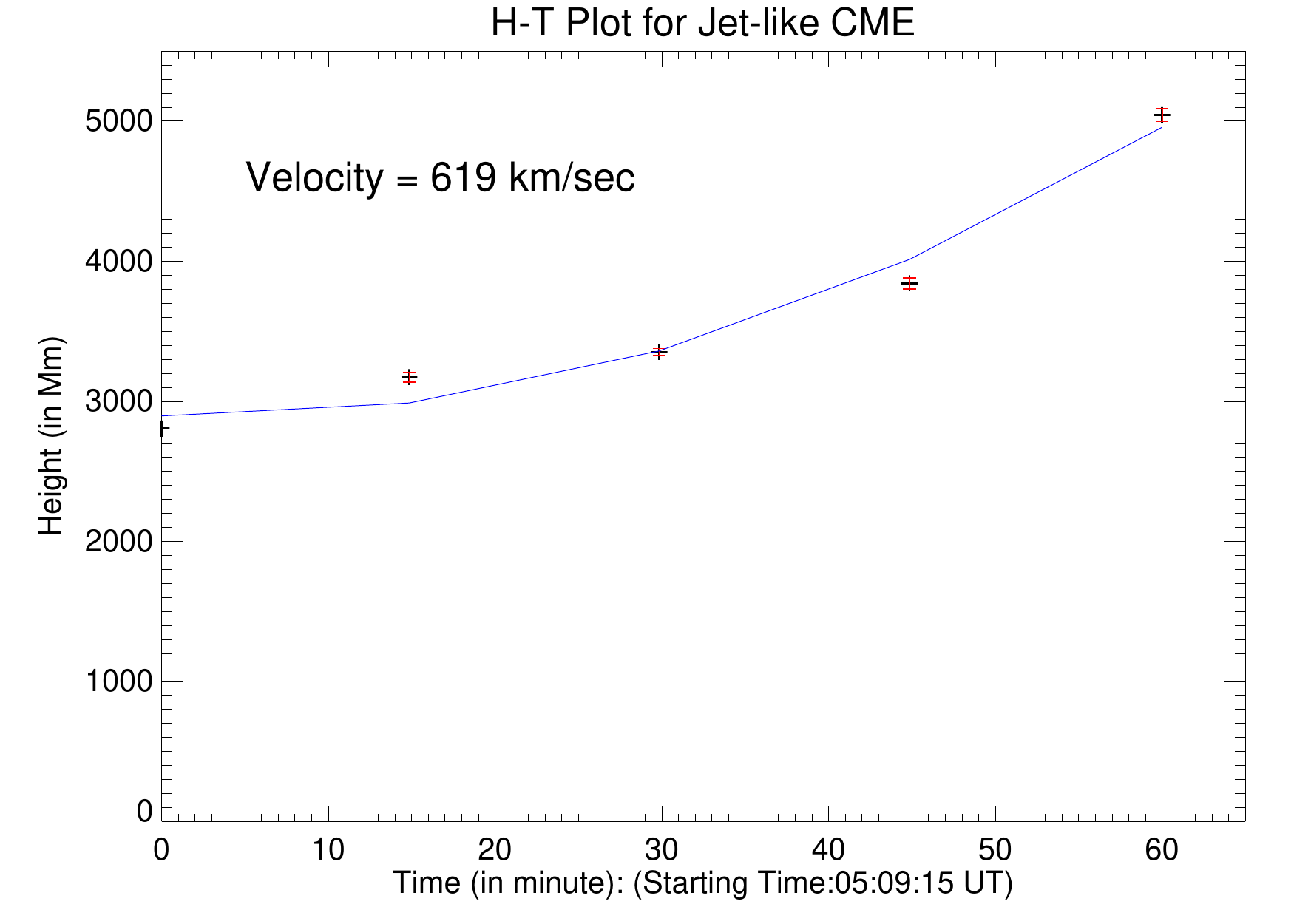}}
\caption{Upper panel shows the tracking of the tip of jet-like CME in STEREO\_A COR2 and STEREO\_B COR2 by using triangulation technique to calculate the projected height of the CME. Bottom panel shows the H-T plot for the jet-like CME.}
\label{Figure 11}
\end{figure*}
\begin{figure*}
\centerline{\includegraphics[height=14cm,width=12cm,keepaspectratio]{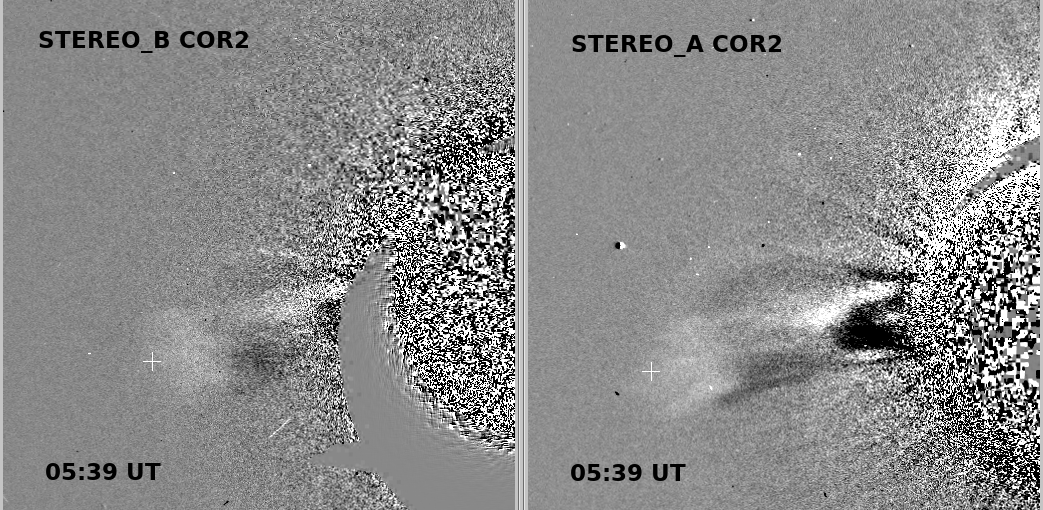}}
\vspace*{0.055\textwidth}
\centerline{\includegraphics[height=12cm,width=11cm,keepaspectratio]{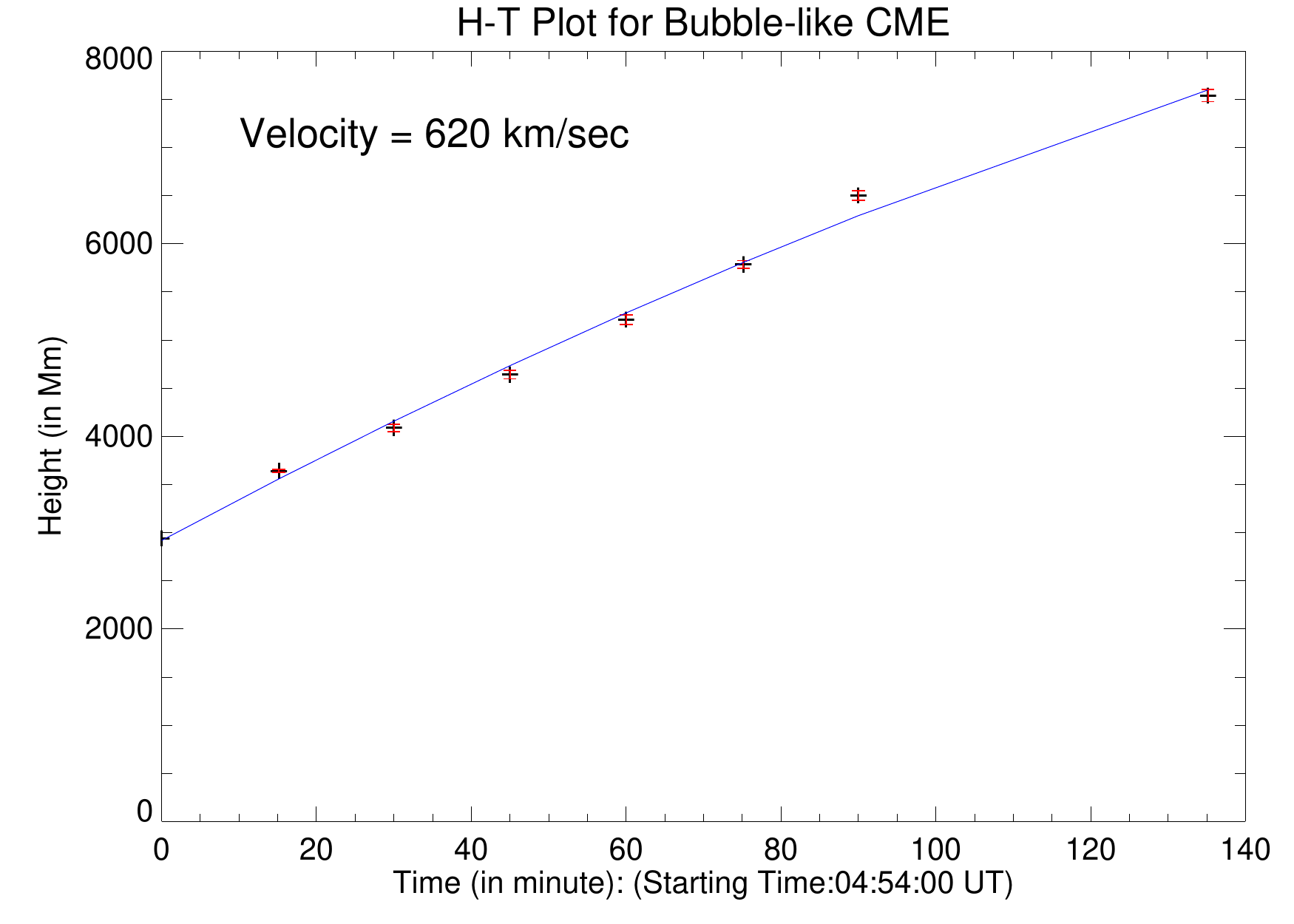}}
\caption{Upper panel shows the tracking of the tip of the bubble-like CME in STEREO\_A COR2 and STEREO\_B COR2 by using triangulation technique to calculate the projected height of the CME. Bottom panel shows the H-T plot for the bubble-like CME.}
\label{Figure 12}
\end{figure*}
\acknowledgments
AKS acknowledges the joint research grant under the frame-work of UKIERI (UK-India Educative and Research Initiatives). 
R.S. thanks the Department of Physics, Indian Institute of Technology (BHU) for providing her Senior Research Fellowship (SRF) and computational facilities. 
We acknowledge the SDO/AIA, SDO/HMI, SoHO/LASCO, STEREO/SECCHI, GONG H$\alpha$ observations for this work. Authors acknowledge Alphonso Sterling, Navdeep Panesar, T. V. Zaqarashvili for their fruitful discussion at initial stage and suggestions. We thank the anonymous referee for his/her valuable comments and suggestions. We thank Sudheer K. Mishra for his help in using tie-pointing method for the kinematics of the blowout jet and twin CME.

\end{document}